\documentclass[twocolumn,aps,amssymb]{revtex4}
\usepackage{graphicx,psfig}
\begin{document}
\title{Wigner Crystallization in mesoscopic 2D electron systems}
\author{A.V.~Filinov$^{1,2}$, M.~Bonitz$^{1}$, and Yu.E.~Lozovik$^2$} 
\address{$^1$Fachbereich Physik, Universit{\"a}t Rostock 
Universit{\"a}tsplatz 3, D-18051 Rostock, Germany \\
$^2$Institute of Spectroscopy, 142090 Troitsk, Moscow Region, Russia} 


\begin{abstract}
Wigner crystallization of electrons in a 2D
quantum dots is reported. It
proceeds in two stages: I) via radial
ordering of electrons on shells and II) freezing of the inter-shell rotation. 
The phase boundary of the crystal is computed
in the whole temperature--density plane, and the
influence of quantum effects and of the particle number is analyzed.
\end{abstract}
\maketitle 


In recent years there is growing interest in {\em finite quantum}
systems at high density or/and low temperature.
In particular, the behavior of
a small number of electrons in quantum dots
is actively investigated both experimentally \cite{ashoori96} and theoretically
\cite{landman99,egger99}. The limiting behavior oftwo-dimensional (2D)
finite quantum systems at {\em zero temperature}
 has been studied by unrestricted Hartree-Fock
calculations \cite{landman99} which revealed a transition from a Fermi liquid
to an ordered state called ``Wigner molecule''. The same
crossover at {\em finite temperature} has been recently demonstrated by
fermionic path integral Monte Carlo \cite{egger99}.It has to be
expected that further increase of correlations (increase of the Brueckner parameter $r_s$) 
will lead to a still higher ordered quantum state resembling the 
Wigner crystal (WC)~\cite{wigner34,wc}. 

On the other hand, for {\em finite classical systems},
Monte Carlo simulations have shown
evidence of crystallization for sufficiently large values of the 
coupling parameter $\Gamma$. These classical clusters consist of well separated shells 
\cite{lozovik87,bedanov94,Schweigert,rakoch98}, and melting proceeds in two stages: 
first, orientational disordering
of shells takes place - neighbouring shells may rotate relative
to each other while retaining their internal order. 
Further growth of thermal
fluctuations leads to shell broadening  and overlap - radial melting. 
The temperature of radial melting $T_r$ may be  many orders of magnitude higher 
than the orientational melting temperature $T_o$ ~\cite{Schweigert}.
Large clusters with $N > 100$ have a regular triangular lattice structure and 
exhibit only radial melting. 

Now the question arises, how does the behavior of finite electron clusters 
change at low temperature, i.e. in the {\em quantum regime}? 
In this Letter we demonstrate that, indeed, Wigner crystallization in 2D 
quantum electron clusters exists and that it is accompanied by two 
distinct - radial and orientiational - ordering transitions too. However, in contrast 
to classical clusters, we observe a new melting scenario which is caused by quantum 
fluctuations and exists even at zero temperature (``cold'' melting \cite{belousov98}).
We present a detailed analysis of the two-stage quantum melting process 
and provide numerical data for the phase boundaries of both crystal 
phases, for particle numbers in the range $N=10\dots 20$.

{\em Model and characteristic parameters.} 
The theoretical analysis of quantum confined electrons at finite temperature
requires the simultaneous account of strong correlations and quantum
effects which excludes e.g. perturbation or mean field methods. We, therefore,
use a path integral Monte Carlo (PIMC) approach.
We consider a finite unpolarized \cite{spin} 2D system of N electrons 
at temperature $T$. The
electrons interact via the repulsive Coulomb potential and are confined in a
harmonic trap of strength $\omega_0$. The system is described by the hamiltonian
\begin{eqnarray}
\hat H &=&\sum\limits_{i=1}^N \frac{\hbar^2 \nabla_i^2}{2 m^*_i} +
\sum\limits_{i=1}^N \frac{m^*_i \omega_0^2 r_i^2}{2} +
\sum\limits^N_{i<j}\frac{e^2}{\epsilon_b |{\bf r}_{i}-{\bf r}_j|},
\label{Hamil}
\end{eqnarray}
\noindent where $m^{*}$ and $\epsilon_b$ are the effective electron mass and background
dielectric constant, respectively. We use the following length and energy scales: 
$r_0$, given by $e^2/\epsilon_b {r_0}=m^*\omega^2 r^2_0/2$,
and $E_c$ - the average Coulomb energy, $E_c = e^2/\epsilon_b {r_0}$.
After the scaling transformations $\{r \rightarrow r/r_0, \;
E \rightarrow E/E_c \}$
the hamiltonian takes the form
\begin{eqnarray}
\hat H &=&  \frac{n^2}{2} \sum\limits_{i=1}^N \nabla_i^2 +
\sum\limits_{i=1}^N r_i^2 +
\sum\limits^N_{i<j}\frac{1}{|{\bf r}_{i}-{\bf r}_j|},
\label{Hamil2}
\end{eqnarray}
\noindent 
where $n$ is the dimensionless density, $n= \sqrt{2} \,l_0^{2}/r^2_0=(a^*_B/r_0)^{1/2}\approx r_s^{-1/2}$, 
$a^*_B$  is the effective Bohr radius, and $l^2_0 = \hbar/(m^{*}\omega_0)$, is
the extension of the ground state wave function of noninteracting trapped 
electrons.(In the crystal phase, $r_0$ is very close to the mean interparticle
distance ${\bar r}$). The dimensionless temperature is given by $T= k_B T/E_c$.

To obtain the configuration and thermodynamic properties of clusters of $N$ electrons 
described by the hamilitonian (\ref{Hamil2}), we performed fermionic PIMC simulations using a standard 
bisection algorithm \cite{Ceperley95}. The number of time-slices 
$M$ has been varied with $n$ and $T$ according to $M=l n/T$, 
where $l$ was typically in the range of $1\dots 10$ to achieve an accuracy better than 
$5 \%$ for the quantities (\ref{angular_dev},\ref{radial_dev}), see below. 
To obtain the phase boundary of the Wigner crystal, calculations 
in a broad range of parameter values $\{n, T, N\}$ were performed. For each 
set of parameters, several independent Monte Carlo runs 
consisting of approximately $10^6$ steps were carried out. 

\begin{figure}[h] 
\centering 
\includegraphics[height=15.5cm]{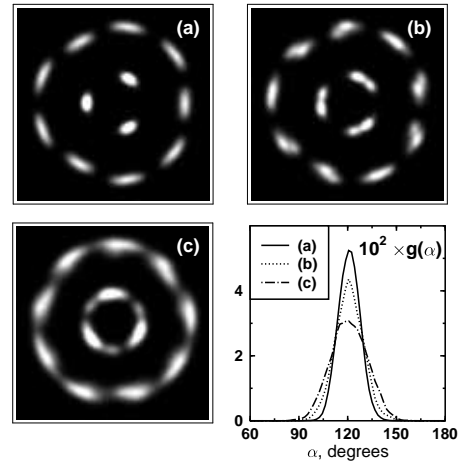} 
\vspace{-8.3cm}
\caption{\small{Electron configurations in the x-y plane for $N=12$, a)  in the orientationally ordered state
 ($n=0.01$); b) at orientational melting ($n=0.014$); c) in the orientationally 
disordered/radially ordered state ($n=0.02$). Lower right figure shows the corresponding angular pair 
distribution  functions for electrons on the inner shell. $T=1.0\times 10^{-4}$.}} 
\label{fig1} 
\end{figure}

\begin{figure}[h] 
\hspace{-1.5cm}\centering 
\includegraphics[height=13cm]{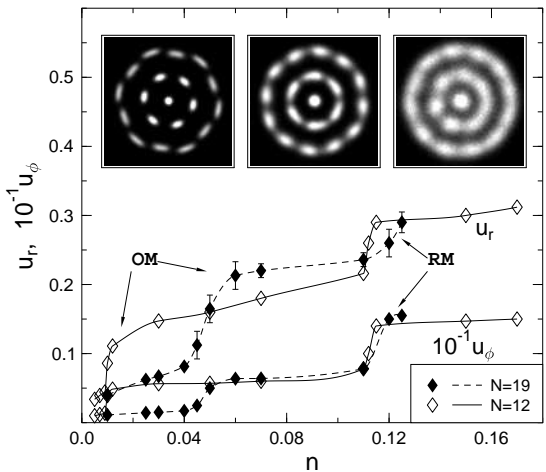} 
\vspace{-5.6cm}
\caption{\small{Relative angular and radial fluctuations, 
Eqs.~(\ref{angular_dev}, \ref{radial_dev}), in the vicinity of the 
orientational (OM) and radial melting (RM) for $N=12$ and $N=19$ versus density. 
Upper three figures show snapshots of the magic cluster with $N=19$ in the three phases (left 
$n=0.025$, middle $n=0.06$, right $n=0.14$). $T=5.0\times 10^{-4}$. Shown error bars are typical for all curves.}} 
\label{fig2} 
\end{figure}

{\em Structure of electron clusters.} The simulations yield the spatial electron configurations
in the trap, examples of which are shown in Figs. \ref{fig1} and \ref{fig2}. One
 clearly sees the  formation of shells. Our analysis revealed the same shell structures 
as reported for the corresponding classical systems \cite{lozovik87,bedanov94}. 
Number of shells and shell occupation depend on $N$, some examples for
 $N=10\dots 20$ are given in Tab.~\ref{tab1}. 
Clusters, in which the particle number on the outer shells are multiples of those 
on the inner shells have the highest symmetry which leads to particular properties. 
Examples of these {\it magic numbers} are $N=10, 12, 16, 19$, cf. Tab.~\ref{tab1}.
Let us now discuss the influence of quantum effects on the clusters.
In contrast to classical systems, where the electrons are point-like 
particles, in our case, the wave function of each electron has a finite width and
may be highly anisotropic, which is typical for low temperature, as is most clearly seen in  
Fig. \ref{fig1}a. This peculiar shape results from a superposition of N-body correlations, 
 quantum effects and the confinement potential. Varying the density and temperature, the shape 
changes in a very broad range, which can lead to qualitative transitions of the cluster, 
including cold quantum melting, as will be shown below.        

{\it Melting and phase transitions.} 
Wigner crystallization is known to occur when the ratio of the Coulomb energy to the kinetic energy,  
$E_c/E_{kin}$, exceeds a certain threshold. In the quantum limit, this ratio is just
the Brueckner parameter $r_s$.
On the other hand, in the classical limit this ratio goes over to 
the classical coupling parameter $\Gamma=E_c/k_B T$. In our units 
$(\bar r \approx r_0)$,
\begin{eqnarray} 
\Gamma=\frac{1}{T} \frac{r_0}{\bar r}\approx \frac{1}{T}.
\end{eqnarray} 
Let us first consider the {\em classical case}. At sufficiently low $T$ (high $\Gamma$),  
a classical cluster is in the orientationally and radially ordered crystal phase. 
Our simulations revealed that in a finite system, (in the classical part of the phase 
space), both orientational and radial melting are determined by characteristic values of $\Gamma$ which, 
however, strongly depend on the particle number $N$.
We find that orientational melting
occurs at temperatures which vary over many orders of magnitude, 
$T^*_o \approx 3\times 10^{-3} \dots 10^{-12}$ and are very sensitive to the shell configuration,  
as observed in Ref. \cite{Schweigert}. 
Higher temperatures (lower $\Gamma$) are found for {\em magic} clusters, cf. Tab.~\ref{tab1}. The 
magic cluster with $N=19$ has unusual stability against inter-shell rotation, as the ratio of 
particle numbers on the three shells is optimal for the formation of a triangular lattice.
In contrast, nonmagic clusters have much lower orientational melting temperatures
(see Tab.~\ref{tab1}), which is particularly striking for $N=20$.

The situation is completely different at radial melting. Here, the critical temperatures 
for inter-shell particle jumps (radial disordering) for magic clusters have been found to be
approximately 2 times smaller
than thosefor nonmagic clusters (Tab.~\ref{tab1}).

We now investigate the phenomenon of {\em quantum orientational melting (OM)}.
As an illustration, consider the cluster with $N=12$ particles.
Fig.~\ref{fig1} shows three snapshots of particle configurations at low temperature 
where the density 
increases from(a) to (c), and (b) corresponds to the critical density. One
of the characteristic features of quantum OM is a high anisotropy of the electron wave
function. Its main spread is along a high and narrow ravine formed
by the many-body potential.

\vspace{-1.5cm}

The growth of quantum fluctuations from (a) 
to (c) is predominantly in angular direction leaving the radial component
practically unchanged.
Fig.~\ref{fig1}(b) shows that, at quantum OM, the probability density of each inner shell electron
splits in two maxima which are due electron transitions between two 
closely lying energy levels. At still higher density,
Fig.~\ref{fig1}(c), the electron wave function spreads further, splitting into 
three maxima, and it starts to spread in radial direction. 
Obviously, OM is
accompanied by intensification of angular distance
fluctuations, cf. Fig.~\ref{fig1}(d). The spread of the electron wave functions smears out the
energy barrier for inter-shell rotation which, in turn, reduces the melting
temperature. In the limit $T\rightarrow 0$ this leads to so-called ``cold'' 
orientational melting.

To allocate the melting temperatures and densities accurately, we examine the relative 
mean angular distance fluctuations of particlesfrom different shells
\begin{eqnarray} 
u_{\phi} \equiv \sqrt{\langle \delta\phi^2 \rangle} =
\frac{2}{ m_{s_1} m_{s_2} } \sum^{m_{s_1}}_{i} \sum^{m_{s_2}}_{j}
\sqrt{\frac{\langle |\phi_j-\phi_i|^2 \rangle}{\langle |\phi_j-\phi_i|\rangle^2} - 1
}, \quad 
\label{angular_dev}
\end{eqnarray}
\noindent where $\phi_i$ and $\phi_j$ are the angular positions of particles
on shells $s_1$ and $s_2$, respectively, and $m_{s_1}$, $m_{s_2}$
are the total number of particles on the shells.
We further consider the magnitude of the relative distance fluctuations
 \begin{eqnarray}
u_{r} \equiv \sqrt{\langle \delta r^2\rangle} = \frac{2}{N(N-1)}\sum^N_{i \le j} \sqrt{
\frac{\langle r_{ij}^2\rangle}{\langle r_{ij}\rangle ^2} - 1 },
\label{radial_dev}
\end{eqnarray}
\noindent where $r_{ij}$ is the distance between particles $i$ and $j$. We found that, in the 
vicinity of orientational and radial melting, quantities (\ref{angular_dev})  and
 (\ref{radial_dev}) show a strong increase, thus providing a suitable quantitative criterion
for these phase transitions, as can be seen in Fig.~\ref{fig2}.
\vspace{-1.7cm}
 
We computed the $n-$dependence of the fluctuations at fixed $T$, as well as the
$T-$dependence at $n=const$ (which corresponds to the dependence on $r_s$ and $\Gamma$, respectively).
The latter was 
mainly used in the classical region of the phase space, where Wigner crystallization occurs at some 
critical value of $\Gamma$, see above.
On the other hand, in the strong quantum limit, there is only a weak temperature dependence, 
so the first method is advantagous. To highlight the peculiarities of quantum melting, 
in the following, we concentrate on the density dependence of the fluctuations .

Fig.~\ref{fig2}  shows the $n$-dependence of the fluctuations
$u_{\phi}$ and $u_{r}$,
at a fixed temperature $T=5\cdot 10^{-4}$, for the magic clusters $N=12$ and
$N=19$.  The first jump  of the fluctuations at
 comparatively low cricitical densities
$n^*_o$ corresponds to quantum OM. At
densities  $n \ge n^*_o$, the angle between particles
from neighbouring shells can take arbitrary values. There is a
significant difference in the magnitude of the melting densities for
the clusters under consideration. As in the classical case, the highest 
stability (highest value $n^*_o$) is found for the magic clusters, with 
the maximum value found for $N=19$.
In particular, we obtained the densities of ``cold'' orientational melting by
extrapolation of our data to zero temperature. The corresponding values for the density and
Brueckner parameter $r^{(o)}_s$ are listed in Tab.~\ref{tab2}.

Let us now proceed in Fig.~\ref{fig2} to higher densities. 
We observe a clear second jump of the fluctuations
$u_{\phi}$ and $u_{r}$, 
which is related to total melting of the WC.
It is instructive to compare the state of the cluster before and after this jump, 
see middle and right shapshots in Fig.~\ref{fig2}, respectively. 
Evidently, when the density is increased, the  peaks of the 
wave functions broaden in radial (and angular) direction until 
their width becomes comparable with the inter-shell spacing. As a result, 
the probability of inter-shell transfer of electrons grows rapidly, causing 
a sudden increase of the radial fluctuations and the onset of radial melting.
Our simulations revealed that the jumps of $u_{\phi}$ and
$u_{r}$ are clearly visible along the whole WC phase
boundary.In the limit of zero temperature, we observe ``cold'' radial melting, 
the corresponding values for the density and $r^{(r)}_s$ are presented in 
Tab.~\ref{tab2}.

\begin{figure}[h] 
\hspace{-3cm}\centering 
\includegraphics[height=5cm]{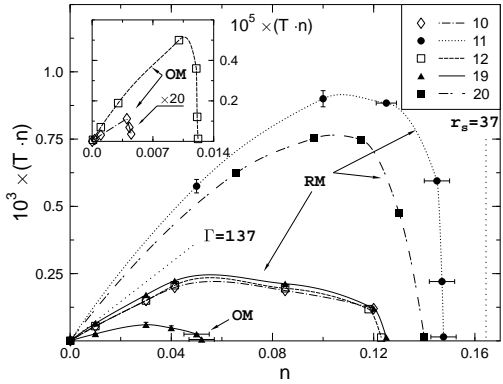} 
\vspace{-1.2cm}
\caption{\small{Phase diagram of the mesoscopic 2D Wigner crystal. ``OM'' (``RM'') denotes 
the orientational (radial) melting curves for 
$N=10,11,12,19,20$. Insert shows an enlarged picture of the 
low-density region. Dotted straight lines indicate the radial melting transition of a macroscopic 
classical and quantum  WC. Brueckner parameter follows from the density by $r_s=1/n^2$.
Shown error bars are typical for
all curves.}} 
\label{fig3} 
\end{figure}

{\em Phase boundary of the mesoscopic Wigner crystal.} 
The obtained values of the melting temperatures and densities 
are now used to draw the phase boundaries of the radially and fully ordered states of mesoscopic 
clusters. The results are summarized in Fig.~\ref{fig3} and Tabs. \ref{tab1}, \ref{tab2} for 
various particle numbers. Consider first the line of radial melting ``RM''. At low densities, 
$n < 0.03$, crystallization occurs at critical values of $\Gamma=1/T^*_r$, see the 
data in Tab.~\ref{tab1}. 
We found that,  for magic (non-magic) clusters, $\Gamma$ is above (below) the 2D bulk value,  
$\Gamma_{\infty}=137$, e.g. \cite{bedanov94}. This shows that non-magic clusters are more stable 
against radial disordering. The same tendency is observed for ``cold'' melting, where the 
critical values of $r_s$ are systematically lower for non-magic clusters. Interestingly, 
for all comparatively small clusters with $10 \le N \le 20$, the critical $r_s$, 
(Tab.~\ref{tab2}), exceeds the known value of an infinite 2D system  
$r_{s}^{\infty} \approx 37$ \cite{tanatar89}. This systematic deviation from the 
critical data of a macroscopic system points to the existence of a different 
scenario of radial melting typical for finite systems. 
For example, for clusters with $N=10$ ($19$), when approaching the phase boundary, we observe 
increasingly frequent transitions between the shell configurations  
$ \{2,8 \} \leftrightarrows \{3,7 \}$ ($\{1,6,12 \} \leftrightarrows \{1,7,11 \}$).
The explanation is the lowering of the  potential barrier 
for inter-shell transitions of individual electrons between these two  
energetically rather close configurations. 

A particularly interesting feature which is missing in infinte systems is the existence of 
the second transition characterized by freezing of the inter-shell rotation, cf. areas bounded 
by the lines ``OM'' in Fig.~\ref{fig3}. This highly ordered state is confined to the region of
low density and low temperature. Its phase boundary is highly sensitive to the angular symmetry
of the cluster. The most symmetric clusters are the magic ones which are essentially more 
stable against orientational melting which is seen in their relatively low values of 
$\Gamma_o^*$ and $r_s^{(o)}$, cf. Tabs. \ref{tab1} and \ref{tab2}. The critical values for the 
non-magic clusters are several orders of magnitude larger, confining their orientationally 
ordered state to much lower densities and temperatures. Finally, the last two columns of 
Tab.~\ref{tab2} contain the maximum temperatures at which the orientationally and 
radially ordered crystal phases may exist.

In summary, we have presented a detailed PIMC-analysis of Wigner crystallization
of finite electron systems confined in a 2D harmonic trap including the critical 
data in the whole density-temperature plane. We have shown that there 
exist two phases characterized by radial and radial plus angular ordering, 
respectively, which is essentially different from macroscopic systems.

\vspace{1cm}

The phase boundaries have been found to be very sensitive to the electron number and to the
shell symmetry. In contrast to classical clusters, we observed ``cold'' orientational 
and radial melting which is governed by the spread of the 
electron wave functions in angular and radial direction. The predictions of our model calculations 
are expected to be relevant in particular for electrons in external fields. Furthermore, 
they lead us to expect Wigner crystallization also in small 2D islands of electrons [holes]
in semiconductor heterostructures: for example, in GaAs/AlGaAs systems, crystallization is 
predicted for carrier densities below approximately
$10^{8}cm^{-2}$ [$(10^{9}\dots 10^{10})cm^{-2}$] and for temperatures below $1.6K \dots 5.5K$, 
for confinement potentials of $3 meV \dots 10meV$. 

This work is supported by the Deutsche Forschungsgemeinschaft
(Schwerpunkt ``Quantenkoh\"arenz in Halbleitern'') and the NIC J\"ulich.

\begin{table}[h]
\caption{\small{{\em Classical} Wigner crystal: critical temperature $T^*$ and coupling 
parameter $\Gamma^*$ corresponding
to the orientational (o) and radial (r) melting for clusters of size $N$ in the
classical limit ($n \rightarrow 0$). Second column contains the shell occupancy, 
starting from the inner shell. Non-magic clusters are typed italic.}}
\label{tab1}
\begin{tabular}{|c|c|c|c|c|c|}
\hline
\hline
$N$ && $T^*_o$ & $\Gamma^*_o$ & $T^*_r$ & $\Gamma^*_r$ \\
\hline
\hline
10  &2,8& $1.5\times 10^{-5}$& $6.7\times 10^{4}$& $5.9\times 10^{-3}$& 169 \\ 
{\it 11} &3,8& ${\it 2.6\times 10^{-7}}$& ${\it 3.85\times 10^{6}}$& ${\it 1.25\times 10^{-2}}$& {\it 80} \\ 
12  &3,9& $8.0\times 10^{-4}$& 1250             & $6.0\times 10^{-3}$& 166 \\ 
16  &1,5, 10& $1.7\times 10^{-3}$& 590             & $6.2\times 10^{-3}$& 161 \\ 
19  &1,6,12&  $ 3.0\times 10^{-3}$& 330 & $6.5\times 10^{-3}$& 154 \\ 
{\it 20} &1,7,12& ${\it 2.9\times 10^{-12}}$& ${\it 3.4\times 10^{11}}$& ${\it 1.2\times 10^{-2}}$& {\it 83} \\ 
\hline
\hline
\end{tabular}
\end{table}

\vspace{-0.5cm}
\begin{table}[h]
\caption{\small{{\em Quantum} Wigner crystal: critical density 
and Brueckner parameter $r_s$ for orientational (o) and radial (r) melting at 
{\em zero temperature}. ${\tilde T}^{max}_{o,(r)}$
is the highest possible melting temperature, ${\tilde T}\equiv n * T.$ The critical parameters 
$n^*_o, r^{(o)}_{s}$ and ${\tilde T}^{max}_o$ for $N=11$ and  $20$ are estimates \protect\cite{ncr}.}}
\label{tab2}
\begin{tabular}{|c|c|c|c|c|c|c|}
\hline
\hline
$N$ & $n^*_o$ & $r^{(o)}_{s}$ & $n^*_r$ & $r^{(r)}_{s}$ & ${\tilde T}^{max}_o$ & ${\tilde T}^{max}_r$  \\
\hline
\hline
10  & $4.5\times 10^{-3}$& $4.9\times 10^{4}$& 0.120 & 69 & $5.0\times 10^{-8}$& $2.2\times 10^{-4}$ \\ 
{\it 11}  & ${\it 5\times 10^{-3}}$ & ${\it 4\times 10^{6}}$ & {\it 0.148} & {\it 45} & ${\it 7\times 10^{-11}}$ & ${\it 9.4\times 10^{-4}}$ \\ 
12  & $1.4\times 10^{-2}$& 5102      & 0.123 & 66 & $5.0\times 10^{-6}$& $2.4\times 10^{-4}$ \\ 
19  & $5.0\times 10^{-2}$& 400            & 0.125 & 64 & $6.0\times 10^{-5}$& $2.5\times 10^{-4}$ \\ 
{\it 20}  & ${\it 1.7\times 10^{-6}}$ & ${\it 3\times 10^{11}}$ & {\it 0.140} & {\it 51} & ${\it 3\times 10^{-18}}$ & ${\it 7.5\times 10^{-4}}$ \\ 
\hline
\hline
\end{tabular}
\end{table}


\begin{references}
\bibitem{wc} We use the notations ``crystal'', ``melting'' and 
``phase transitions'' to underline the analogy to the corresponding 
phenomena in infinite systems.
\bibitem{spin} The WC is very weakly influenced by spin correlations which
become important at strong overlap of individual 
electrons at $r_s<10$ \cite{egger99}, cf. Tab.\ref{tab2}.
\bibitem{ncr} The simple relations
$n_{o,r}^*\sim \sqrt{T_{o,r}^*}$ 
and ${\tilde T}_{o,r}^{max} \sim n_{o,r}^*/(2 \Gamma_{o,r}^*)$ are in good 
qualitative agreement with our simulation data.
\end{references}
\end{document}